\documentclass[twocolumn]{aastex63}
\usepackage{graphicx}
\usepackage{amsmath}
\usepackage{apjfonts,natbib}
\usepackage{appendix}
\usepackage{ulem}

\newcommand{\ha}{H\ensuremath{\alpha}}
\newcommand{\pa}{Pa\ensuremath{\alpha}}
\newcommand{\hb}{H\ensuremath{\beta}}

\newcommand{\mgii}{Mg\,{\footnotesize II}}

\newcommand{\lbol}{\ensuremath{L\mathrm{_{bol}}}}

\newcommand{\msun}{\ensuremath{M_{\odot}}}

\newcommand{\mbh}{\ensuremath{M_\mathrm{BH}}}

\newcommand{\cc}{\hbox{cm$^{-3}$}}

\newcommand{\cms}{\hbox{cm$^{-2}$s$^{-1}$}}
\newcommand{\cmii}{\hbox{cm$^{-2}$}}
\newcommand{\ergcms}{\ifmmode {\rm ergs\,cm}^{-2}\,{\rm s}^{-1} \else ergs\,cm$^{-2}$\,s$^{-1}$\fi}
\newcommand{\ergcmsA}{\ifmmode{\rm ergs}\, {\rm cm}^{-2}\,{\rm s}^{-1}\,{\rm\AA}^{-1} \else ergs\, cm$^{-2}$\, s$^{-1}$\, \AA$^{-1}$\fi}
\newcommand{\ergcmsHz}{\ifmmode{\rm ergs\,cm}^{-2}\,{\rm s}^{-1}\,{\rm Hz}^{-1} \else ergs\,cm$^{-2}$\,s$^{-1}$\,Hz$^{-1}$\fi}
\newcommand{\phcms}{\ifmmode {\rm ph\,cm}^{-2}\,{\rm s}^{-1} \else ,ph\,cm$^{-2}$\,s$^{-1}$\fi}
\newcommand{\phcmsA}{\ifmmode {\rm ph\,cm}^{-2}\,{\rm s}^{-1}\,{\rm\AA}^{-1} \else ph\,cm$^{-2}$\,s$^{-1}$\,\AA$^{-1}$\fi}

\shorttitle{BLR sizes of RM and SA}
\shortauthors{Zhang et al.}
\begin{document}
	
\title{The Deviation of the Broad-line Region Size Between Reverberation Mapping and Spectroastrometry}
\correspondingauthor{Zhicheng He}
\email{zcho@ustc.edu.cn}

\author{Xiaer Zhang}
\affiliation{CAS Key Laboratory for Research in Galaxies and Cosmology, Department of Astronomy, University of Science and Technology of China, Hefei, Anhui 230026, China}
\affiliation{School of Astronomy and Space Science, University of Science and Technology of China, Hefei, Anhui 230026, China}

\author[0000-0003-3667-1060]{Zhicheng He}
\affiliation{CAS Key Laboratory for Research in Galaxies and Cosmology, Department of Astronomy, University of Science and Technology of China, Hefei, Anhui 230026, China}
\affiliation{School of Astronomy and Space Science, University of Science and Technology of China, Hefei, Anhui 230026, China}

\author{Tinggui Wang}
\affiliation{CAS Key Laboratory for Research in Galaxies and Cosmology, Department of Astronomy, University of Science and Technology of China, Hefei, Anhui 230026, China}
\affiliation{School of Astronomy and Space Science, University of Science and Technology of China, Hefei, Anhui 230026, China}

\author[0000-0001-8416-7059]{Hengxiao Guo}
\affiliation{Department of Physics and Astronomy, 4129 Frederick Reines Hall, University of California, Irvine, CA, 92697-4575, USA}

\begin{abstract}
The combination of the linear size from reverberation mapping (RM) and the angular distance of the broad 
line region (BLR) from spectroastrometry (SA) in active galactic nuclei (AGNs) can be used to measure the Hubble 
constant $H_0$. Recently, \cite{wang2020} successfully employed this
approach and estimated $H_0$ from 3C 273. However, there may be a systematic deviation between the
response-weighted radius (RM measurement) and luminosity-weighted radius (SA measurement), especially
when different broad lines are adopted for size indicators (e.g., \hb\ for RM and \pa\ for SA). 
Here we evaluate the size deviations measured by six pairs of hydrogen lines (e.g., \hb, \ha\ and \pa) via the locally 
optimally emitting cloud (LOC) models of BLR. We find that the radius ratios $K$(=$R_{\rm SA}$/$R_{\rm RM}$) of the same line 
deviated systematically from 1 (0.85-0.88) with dispersions between 
0.063-0.083. Surprisingly, the $K$ values from the \pa(SA)/\hb(RM) and \ha(SA)/\hb(RM) pairs not only are closest 
to 1 but also have considerably smaller uncertainty. 
Considering the current infrared interferometry technology, the \pa(SA)/\hb(RM) pair is the ideal 
choice for the low redshift objects in the SARM project. In the future, the \ha(SA)/\hb(RM) pair
could be used for the high redshift luminous quasars. These theoretical estimations 
of the SA/RM radius pave the way for the future SARM measurements to further constrain the standard 
cosmological model.
\end{abstract}


\section{Introduction}
The Hubble constant $ H_0$ is a fundamental parameter of the precision cosmology. However, there is a significant difference (up to $4.4\sigma$) between the Planck measurement from cosmic microwave background (CMB) anisotropies, and supernovae Ia (SNIa) measurements calibrated with Cepheid distances (\citealt{freedman2017,riess2019,Planck2018}). Therefore, a better and independent measurement of $ H_0$ tension is urgently needed.  

Active galactic nuclei (AGNs) are ubiquitous and the most luminous persistent celestial objects in the universe. They have the potentiality to establish as cosmological probes based on some of their features, such as the nonlinear relation between the UV and X-ray luminosities (\citealt{risaliti2019}), the short-term UV/optical variability amplitude (\citealt{sun2018}), the wavelength-dependent time delays of continuum flux variations \citep{collier1999, cackett2007} and the lag-luminosity relationship for Broad Line Region (BLR) (\citealt{watson2011,czerny2013}) or 
the dusty torus (\citealt{honig2011,honig2014,koshida2014,honig2017,he2021}).

\cite{elvis2002} proposed a pure geometrical method to determine the distance to quasars with $D_{\rm A} = R_{\rm BLR}/\theta$, where $R_{\rm BLR} = c\tau$, $\tau$ is the light-travel time from the centre to the BLR, and $\theta$ is the resolved angular size of the BLR from interferometric 
or the technique of spectroastrometry (SA) \citep{bailey1998}. Note that the SA method resolves the structure of the target source by measuring the wavelength dependence of the position of an object. It has the important advantage that it can, in principle, provide information 
on the spatial structure of the object on scales much smaller than the diffraction limit of the telescope being used \citep{bailey1998}.
Compared with other tools to measure the cosmological distances, it has three primary advantages: 1) it is a model-independent method; 2) its uncertainties could be reduced by repeating observations of a medium-size AGN sample \citep{wang2020, songsheng2021}; 
3) AGNs are luminous and scattered in all directions over a wide range of redshift, which can be used to test the potential anisotropy of the accelerating expansion of the Universe. 

Recently, GRAVITY at the very large telescope interferometer (VLTI) successfully revealed the structure, kinematics and angular sizes of the 
BLR of 3C 273, an AGN at $z=0.15834$, using the SA method \citep{abuter2017, collaboration2018}. \cite{wang2020} firstly combined the SA measurement and the reverberation mapping (RM) measurement, namely SARM project. Based on the SARM analysis, \cite{wang2020} are able to determine the angular distance of $551.5^{+97.3}_{-78.7}$ Mpc to 3C 273, thus to constrain $ H_0$ = $71.5^{+11.9}_{-10.6}$ km s$^{-1}$ Mpc$^{-1}$ with the $z-D_{A}$ 
relation \citep{peacock1998}. 

However, RM radius and SA radius could be different things: RM radius represents the variable components of the BLR region and is known as the response-weighted (or flux variation weighted) radius, while the SA radius represents the flux-weighted region. Furthermore, the RM and SA measurements adopted two different lines: \pa\ for SA and \hb\ for RM measurements due to its difficulty of reverberation for an infrared emission line with relative weaker variability and longer lags w.r.t. optical lines. 
Consequently, there may be a systematic deviation between these two type radius.

In this work, we will systematically evaluate this probable deviation of SA/RM radius for several observable hydrogen 
lines (e.g., \hb, \ha, and \pa) in the future SARM project 
based on the locally optimally emitting cloud (LOC) model of BLR \citep{baldwin1995}. The paper is organized as follows. In Section \ref{sec:simu}, we illustrate the set up of the LOC simulations. The difference of the SA and RM sizes is calculated for several hydrogen lines in Section \ref{sec:result}. We conclude in Section \ref{sec:con}.

\section{Photoionization Simulation}\label{sec:simu}
\subsection{LOC model}
We use the photoionization code CLOUDY17.01 (\citealt{ferland2017}) to carry out the simulations for the line emission of BLR
based on the LOC model which is a physically motivated photoionization model for the BLR \citep{baldwin1995}.
In the LOC model, the BLR consists of clouds with different gas densities and distances from the central continuum source with 
an axisymmetric distribution. The total emission line intensity we observe originates from the combination of all clouds but is dominated by those with the highest efficiency of reprocessing the incident ionizing continuum.
As a result, the total emission line intensity can be calculated by the formula below (\citealt{baldwin1995}):
\begin{equation}
L_{\rm line} \propto \iint_{R_{\rm in}}^{R_{\rm out}} r^{2}F(r)f(r)g(n)dndr,
\label{eq1}
\end{equation}
where $F(r)$ is the emission intensity of a single cloud at the radius $r$, $f(r)$ is the cloud covering factors, and $g(n)$ is the cloud distribution function. 
The distribution functions of clouds, i.e., $f(r)$ and $g(n)$, can be specified by the observed emission-line properties. According to \cite{baldwin1995}, $f(r)$ and $g(n)$ can be simplified as power-law functions: 
$f(r) \propto r^{\Gamma} $ and $g(n) \propto n^{\beta}$ with $\Gamma =-1$ and $\beta =-1$. 
For the best-known NGC 5548, \cite{korista2000} constrained $ -1.4 <\Gamma < -1$ by the observed time-averaged UV spectrum.
Based on a large sample of 5344 quasar spectra taken from the SDSS Data Release 2, the parameters $\Gamma$ and $\beta$ are constrained to
$\Gamma =-1.52\pm 0.13$ and $\beta =-1.08\pm 0.05$ \citep{nagao2006}. Here, we adopt a sufficiently wide parameter range of $\Gamma$ from $-1.5$ to $-0.5$, and a fixed $\beta =-1$ (since previous studies, e.g., \citet{korista2000}, suggest that $\beta$ should not to be far from $-1$) to evaluate the deviation of SA/RM radius for several observable hydrogen lines.

We assume a typical AGN with the black hole mass $\mbh = 10^{8} \msun$ 
and bolometric luminosity $\lbol=10^{45}\rm erg\ s^{-1}$ w.r.t. the Eddington ratio $L_{\rm bol}/L_{\rm edd}\simeq 0.1$ in our simulation. Considering the generality, we use a typical radio-quiet AGN SED \citep{dunn2010} for the incident SED, resulting in an emission rate of hydrogen-ionizing photons of $Q_{\rm H} =10^{55}\  \rm s^{-1}$.
The outer BLR boundary determined by dust sublimation of the inner edge of the torus corresponding to a surface ionizing flux $ \log_{10}\Phi (\rm H)=17.9 cm^{-2}s^{-1}$
\citep{nenkova2008,landt2019}, which is about 140 lt-day in the best-studied AGN NGC 5548 (e.g., \citealt{korista2000}), with an average $Q_{\rm H} =10^{54.13}\  \rm s^{-1}$ for the BLR.
According to the definition of surface ionizing flux:
\begin{equation}
\Phi_{\rm H}=\frac{Q_{\rm H}}{4\pi R^2},
\label{eq2}
\end{equation}
the outer BLR boundary is $R_{\rm out}=10^{18}\ \rm  cm$ for $Q_{\rm H} =10^{55}\  \rm s^{-1}$. 
Furthermore, both LOC calculation of the most extended \mgii\ line \citep{guo2020} and near-IR reverberation measurements \citep{kishimoto2007} constrained that the outer BLR boundary and the innermost torus radius are both a factor of $\sim$3 less than or approximately equal to $ 10^{18}\ \rm  cm$. Therefore, we assume the range of outer BLR boundary is $R_{\rm out}=(10^{17.5}- 10^{18})\ \rm  cm$ to evaluate the deviation of SA/RM radius.

In our calculation, we set the inner BLR boundary to be two orders of magnitude smaller than the outer BLR boundary (\citealt{landt2014}), i.e.,  $R_{\rm in}=10^{16}\ \rm  cm$.
\cite{guo2020} showed that the influence of the uncertainty of $R_{\rm in}$ on the final result can be negligible.
We consider the range of the gas number density: $10^{8} \cc \le n_{\rm H} \le 10^{12}\cc$, since below $n_{\rm H}=10^{8} \cc$ the clouds are inefficient in producing 
emission lines and above $n_{\rm H}=10^{12} \cc$ the clouds mostly produce thermalized continuum emission rather than emission lines \citep{korista2000}.
In addition, we adopt the metallicity $Z = Z\sun$ in our calculation. Note that, since we only focus on the hydrogen emission lines, the metallicity will not affect our calculations. The overall covering factor is set to CF = 50\%, as adopted in \cite{korista2004}. 

As shown in Figure 1, we calculate the \ha, \hb, and \pa\ emissions at different gas densities $n_{\rm H}$ and surface ionizing fluxes $\Phi(\rm H)$
for a cloud with a hydrogen column density $N_{\rm H}=10^{23}\cmii$. For a fixed $Q_{\rm H}$, the $\Phi(\rm H)$ is inversely proportional to the square of the distance,
i.e., $\Phi(\rm H)$ $\propto 1/r^{2}$.  For $Q_{\rm H} =10^{55}\  \rm s^{-1}$, $R_{\rm in}=10^{16}\ \rm  cm$ and $R_{\rm out}=10^{18}\ \rm  cm$ correspond to $\Phi(\rm H)=10^{21.9}\cms$ and $10^{17.9}\cms$, respectively.
So, we integrate emission intensity to generate the total flux of \ha, \hb, and \pa\ in the ranges of $\Phi(\rm H)$ from $10^{17.9}\cms$ to $10^{21.9}\cms$ and $n_{\rm H}$ from $10^{8} \cc$ to $10^{12} \cc$.

\begin{figure*}
\centering
\includegraphics[width=19.0cm]{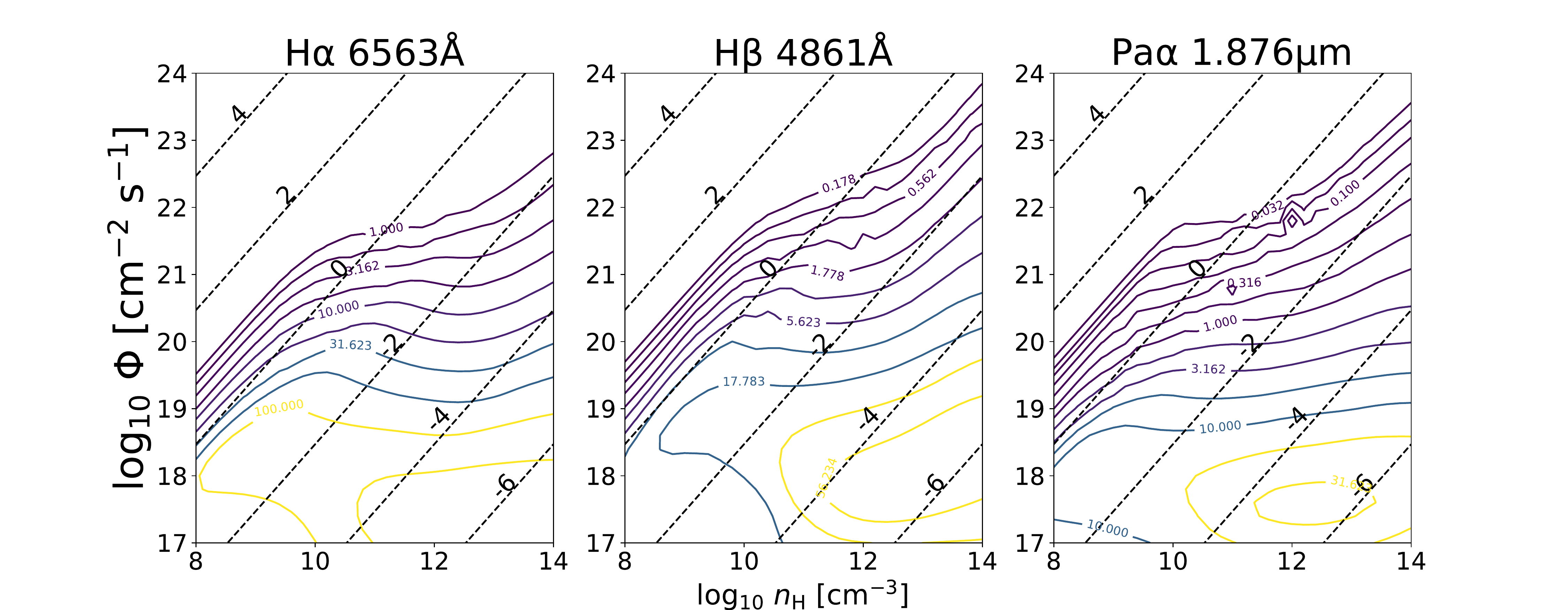}
\caption{The equivalent width (EW) of  \ha, \hb, and \pa\ emissions at different gas densities $n_{\rm H}$ and surface ionizing fluxes $\Phi_{\rm H}$
for a cloud with a hydrogen column density $N_{\rm H}=10^{23}\cmii$. The value of EW is marked for each contour.
The dashed diagonal lines are photoionization parameters decreasing from the upper left ($\log_{10}U = 4$) to the lower right ($\log_{10}U = -6$).}
\end{figure*}

\begin{figure*}
\centering
\includegraphics[width=8.cm]{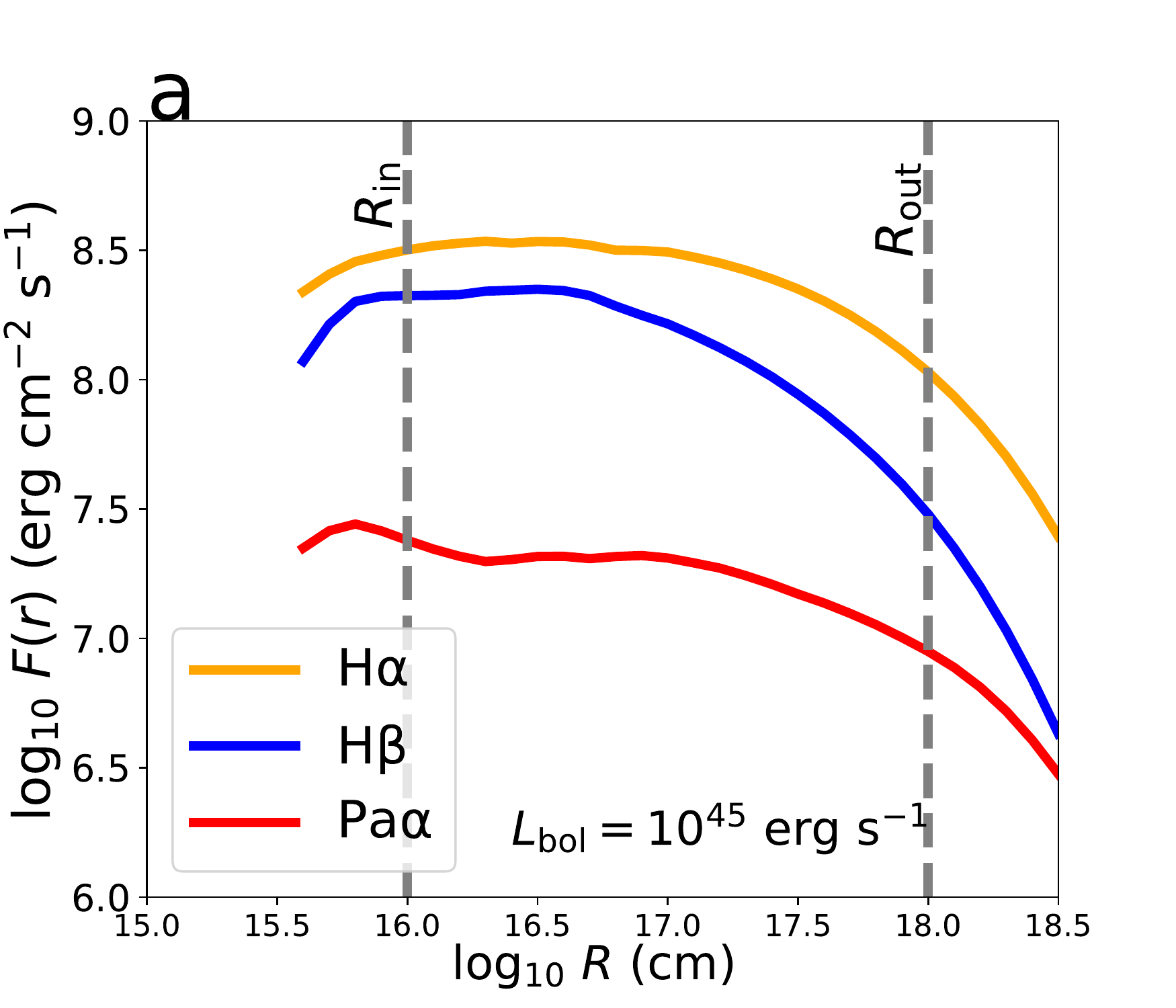}
\includegraphics[width=8.cm]{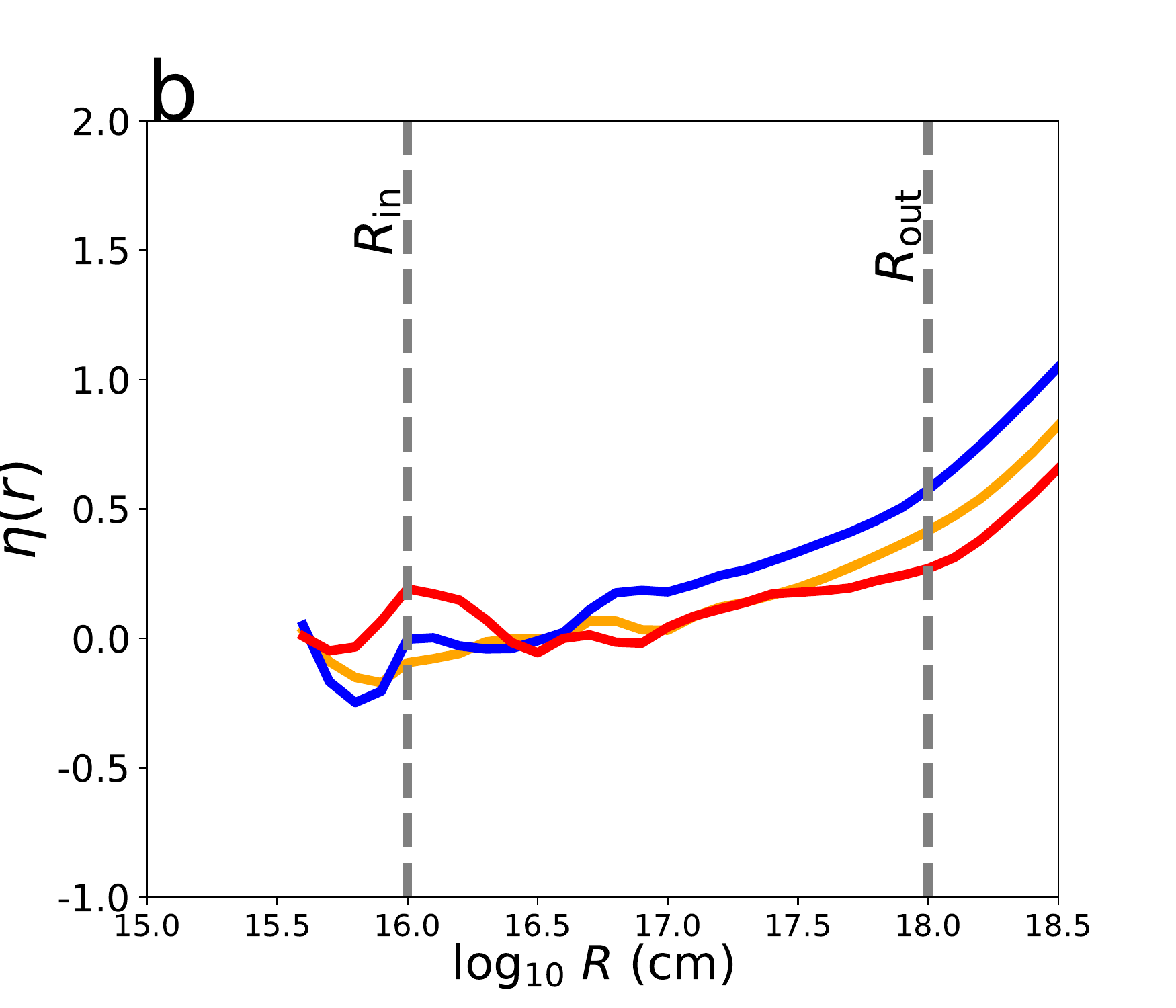}
\caption{Panel a: radial emissivity function $F(r)$. Panel b: radial responsivity function $\eta(r)$
 is calculated based on Equation (4). $R_{\rm in}$ and $R_{\rm out}$ are the inner and outer BLR boundaries in our calculation. }
\end{figure*}

\subsection{The definition of SA and RM size}
The spectroastrometry size of BLR is actually the flux-weighted radius. As a result, the SA radius can be calculated as follows:
\begin{equation}
R_{\rm SA}= \frac{\iint_{R_{\rm in}}^{R_{\rm out}} r^{3}F(r)f(r)g(n)dndr}{\iint_{R_{\rm in}}^{R_{\rm out}} r^{2}F(r)f(r)g(n)dndr}.
\label{eq3}
\end{equation}
The $F(r)$ curves of \ha, \hb, \pa\ are shown in Figure 2\textbf{a}.

The reverberation mapping (RM) size is actually the response-weighted radius. 
According to the previous works \citep{goad1993, korista2004, goad2014}, 
the emission-line responsivity (in logarithmic space) to the changes in the incident hydrogen ionizing photon flux, can be written as:
\begin{equation}
\eta(r)=\frac{d\log_{10} F(r)}{d\log_{10} \Phi_{\rm H}}\propto -0.5\frac{d\log_{10} F(r)}{d\log_{10} r}, {\rm since}\  \Phi_{\rm H} \propto r^{-2}.
\label{eq3}
\end{equation}
The $\eta(r)$ of \ha, \hb, \pa\ are shown in Figure 2\textbf{b}.
In linear space, the emission-line responsivity $dF(r)$/ $d\Phi_{\rm H}$=$[F(r)/\Phi_{\rm H}] \eta(r) \propto F(r)\eta(r)r^2$.
As a result, the RM radius can be calculated as follows:
\begin{eqnarray}
R_{\rm RM}&=& \frac{\iint_{R_{\rm in}}^{R_{\rm out}} r^{3}[dF(r)/d\Phi_{\rm H}]f(r)g(n)dndr}{\iint_{R_{\rm in}}^{R_{\rm out}} r^{2}[dF(r)/d\Phi_{\rm H}]f(r)g(n)dndr} \\
&=& \frac{\iint_{R_{\rm in}}^{R_{\rm out}} r^{5}F(r)\eta(r)f(r)g(n)dndr}{\iint_{R_{\rm in}}^{R_{\rm out}} r^{4}F(r)\eta(r)f(r)g(n)dndr}.
\label{eq4}
\end{eqnarray}

\begin{figure*}
	\centering
	\includegraphics[width=19cm]{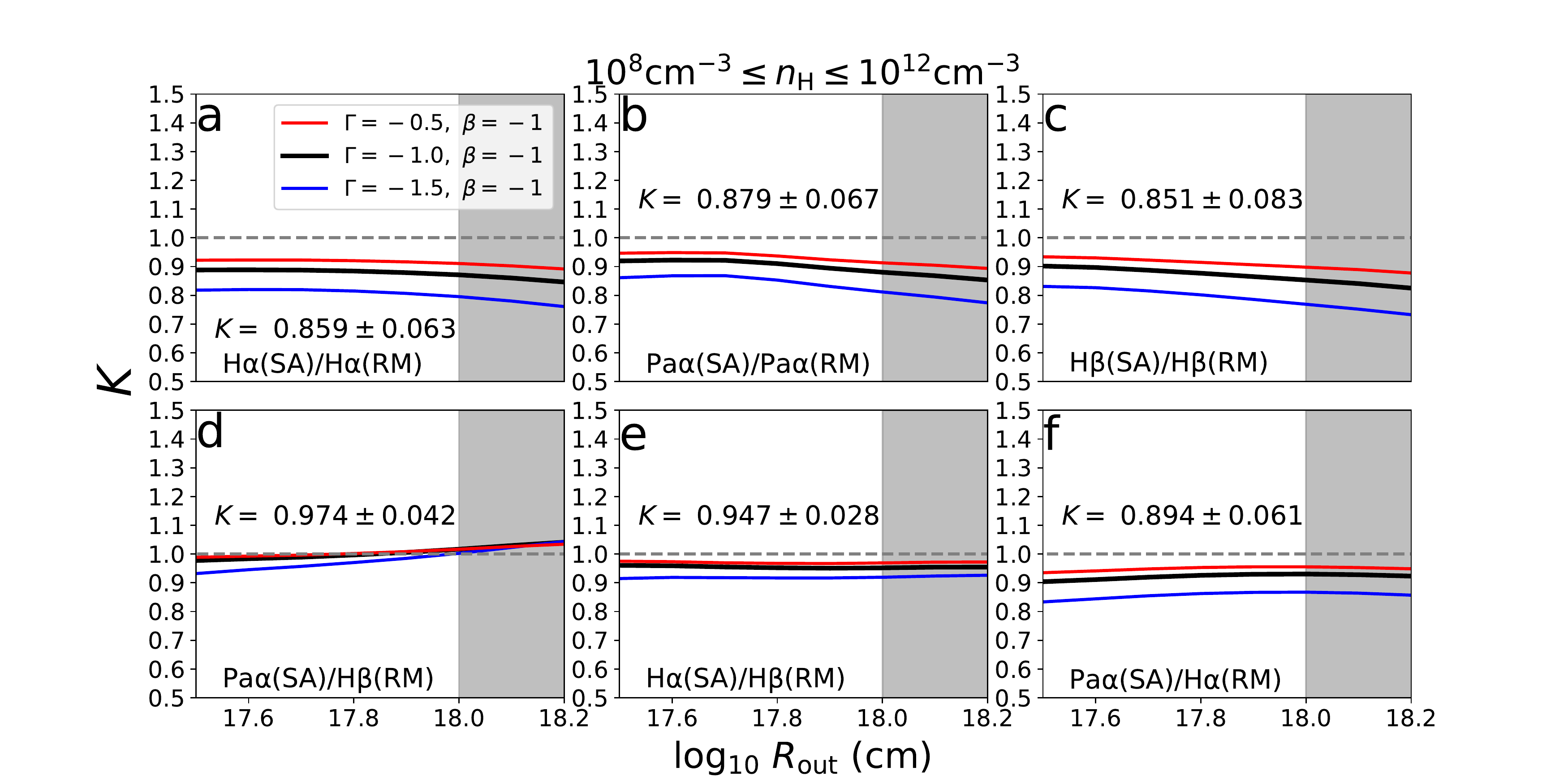}
	\caption{The radius ratio $K$ between of SA and RM size for several main hydrogen lines. 
Panels \textbf{a, b, c}: the combination of the same lines. Panels \textbf{d, e, f}: the combination of the different lines. 
The gray shadow is the region beyond the dust-limited outer BLR boundary $R_{\rm out}=10^{18}\ \rm  cm$.
We estimate the uncertainty of radius ratio $K=R_{\rm SA}/R_{\rm RM}$ under wide range of parameters:
$ -1.5 <\Gamma < -0.5$, $\beta =-1$ and $10^{17.5} {\rm cm} <R_{\rm out}<10^{18} {\rm cm}$.
The average and half-range of $K$ is marked in each panel.
Surprisingly, not only the uncertainty of $K$ from the combination of the different lines are less than that from 
the combination of the same lines, but also the average values of $K$ from the combination of the different lines are closer to 1. }
\end{figure*}

\begin{figure*}
	\centering
	\includegraphics[width=8cm]{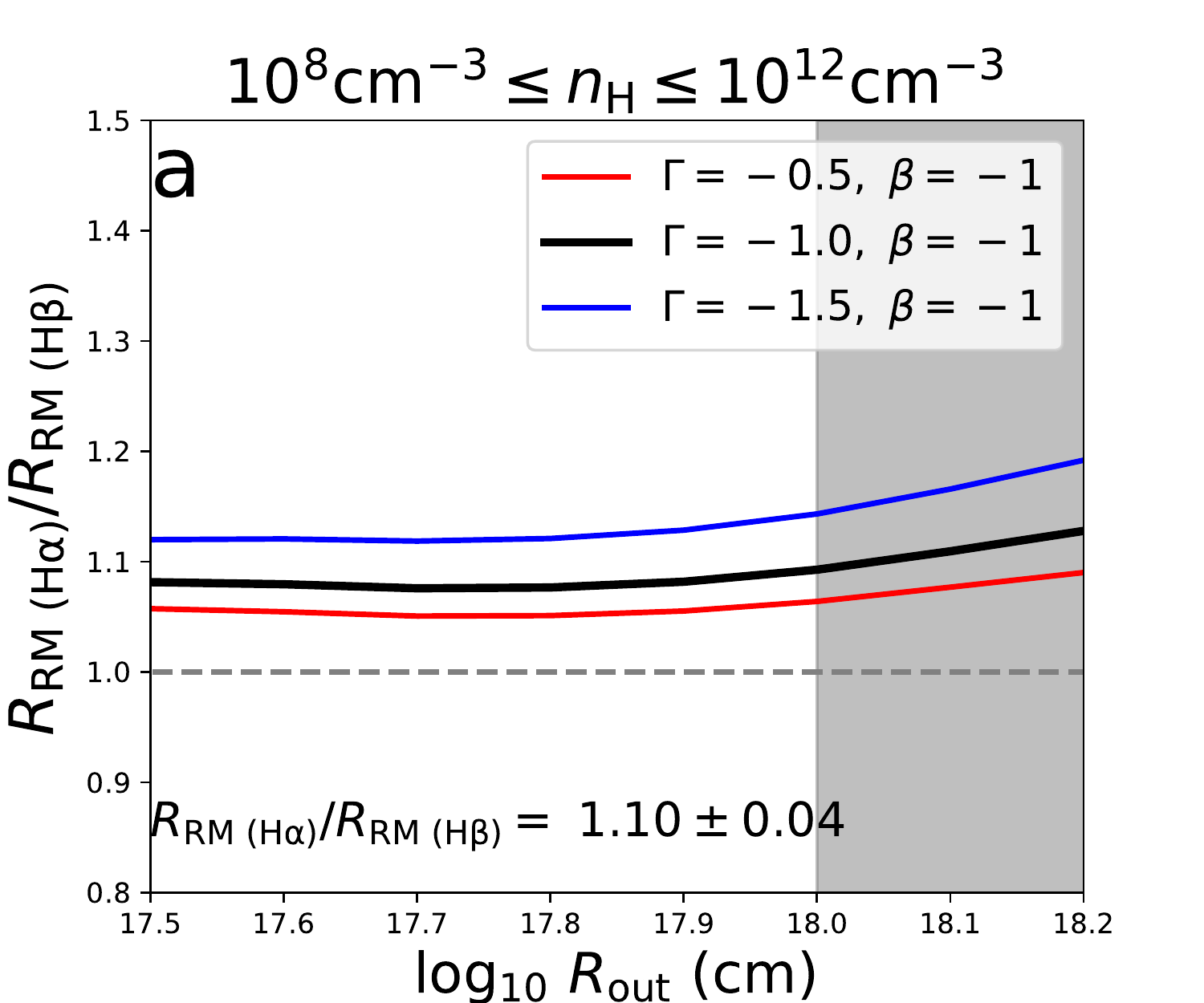}
	\includegraphics[width=8cm]{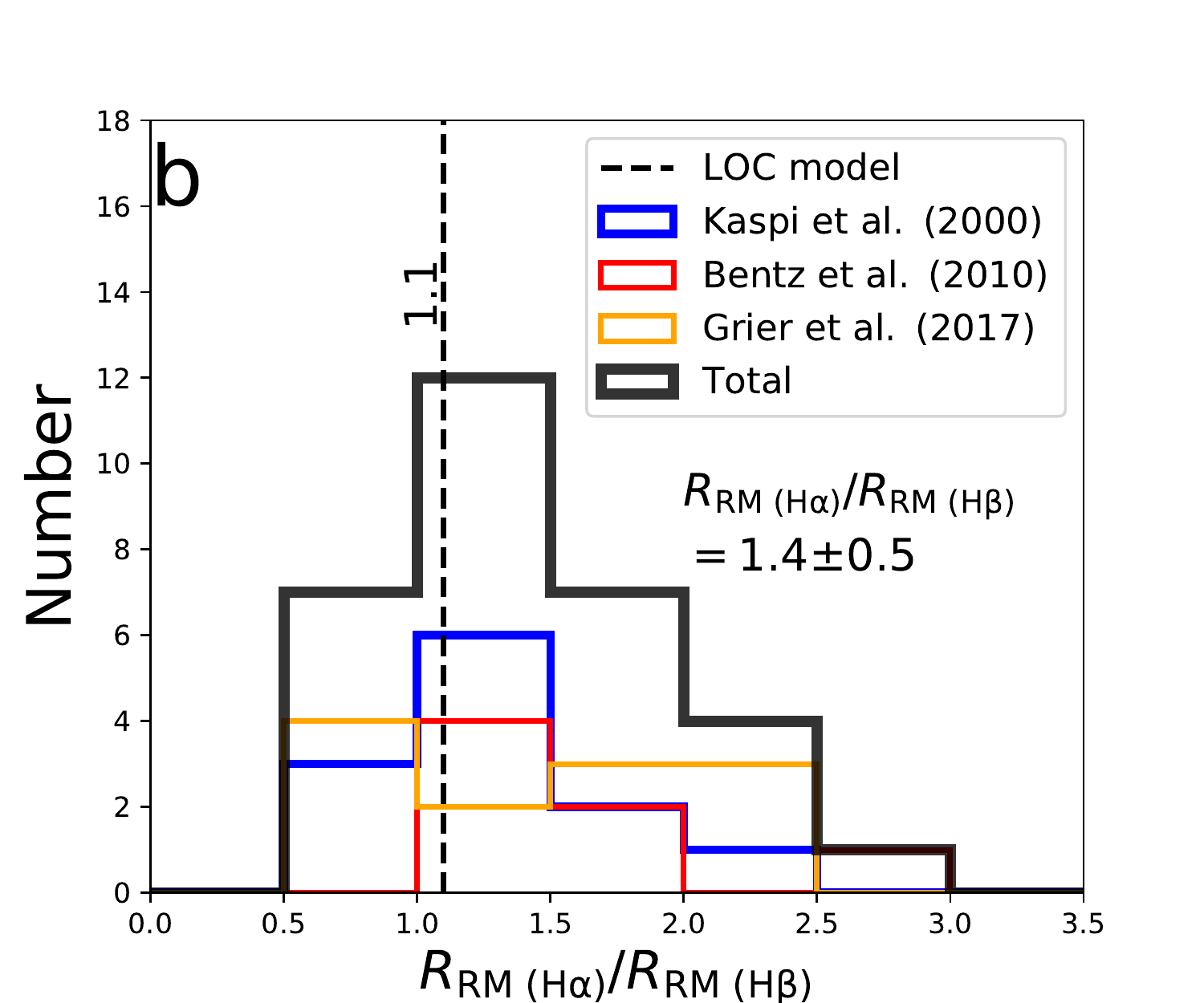}
	\caption{ Panel \textbf{a}: the simulation radius ratio of \ha\ to \hb\ of the RM measurements is 
$R_{\rm RM\ (\ha)}/R_{\rm RM\ (\hb)}=1.10\pm 0.04$. All the symbols are the same as Figure 3.
Panel \textbf{b}: the observational results of $R_{\rm RM\ (\ha)}/R_{\rm RM\ (\hb)}$.
The blue histogram represents measurements from \citet{Kaspi2000}. The red histogram represents measurements from \citet{Bentz2010}. 
The orange histogram represents measurements from \citet{Grier2017}. The black histogram represents the sum of the above three samples.
The mean and standard deviation of the distribution of observational $R_{\rm RM\ (\ha)}/R_{\rm RM\ (\hb)}$ are $1.4\pm 0.5$.
The vertical dashed line marks the mean value of simulated $R_{\rm RM\ (\ha)}/R_{\rm RM\ (\hb)}$ based on LOC model.
This simulation result is roughly consistent with the observational results.}
\end{figure*}

\begin{figure*}
	\centering
	\includegraphics[width=19cm]{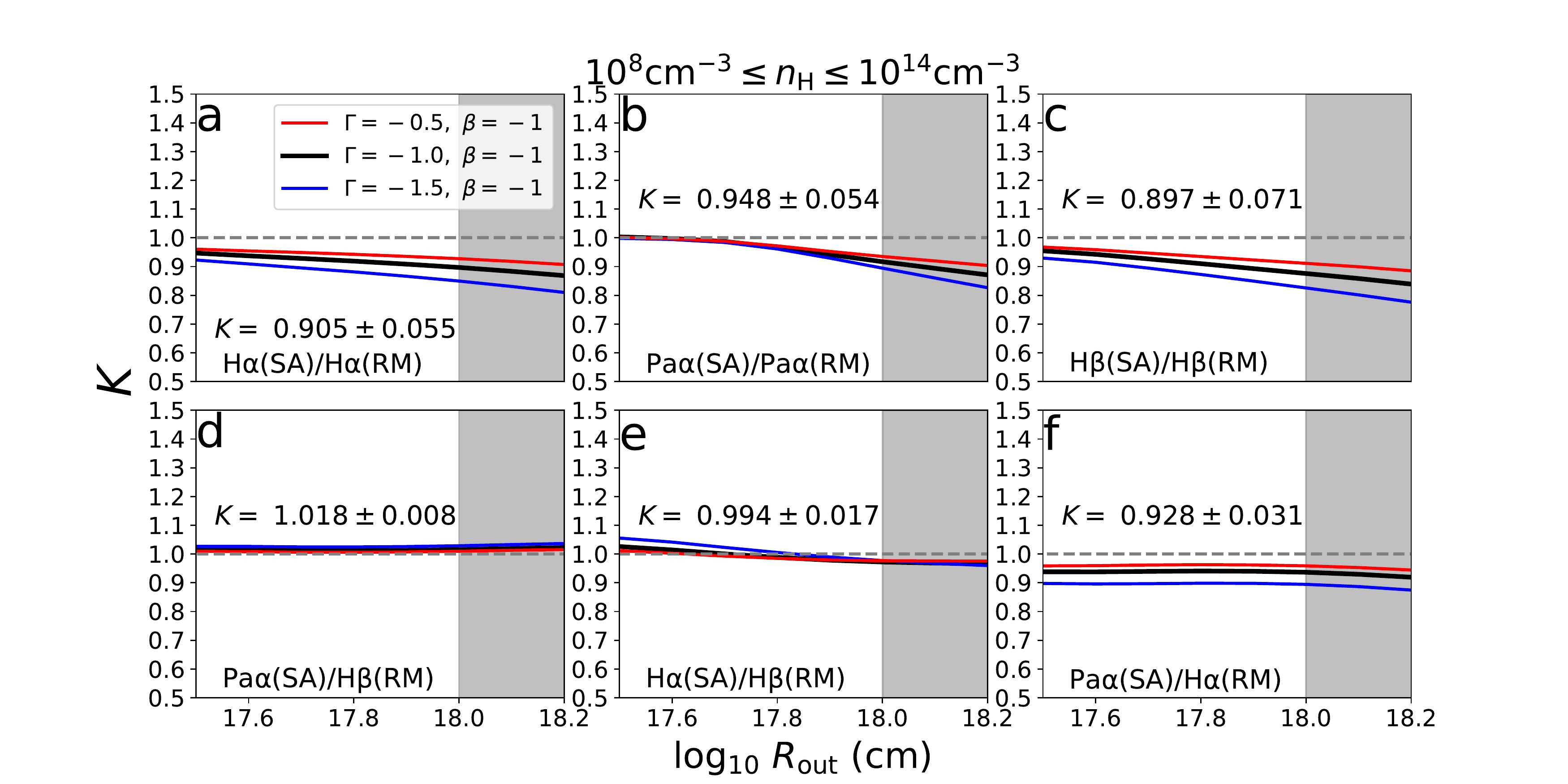}
	\caption{The radius ratio $K$ between of SA and RM size for several main hydrogen lines. All the parameters ranges and
symbols are the same as Figure 3, except the gas density range. Here the density range is 
$10^8{\rm cm^{-3}} \leq n_{\rm H} \leq 10^{14} \rm cm^{-3}$. 
 }
\end{figure*}

\begin{figure}
	\centering
	\includegraphics[width=9cm]{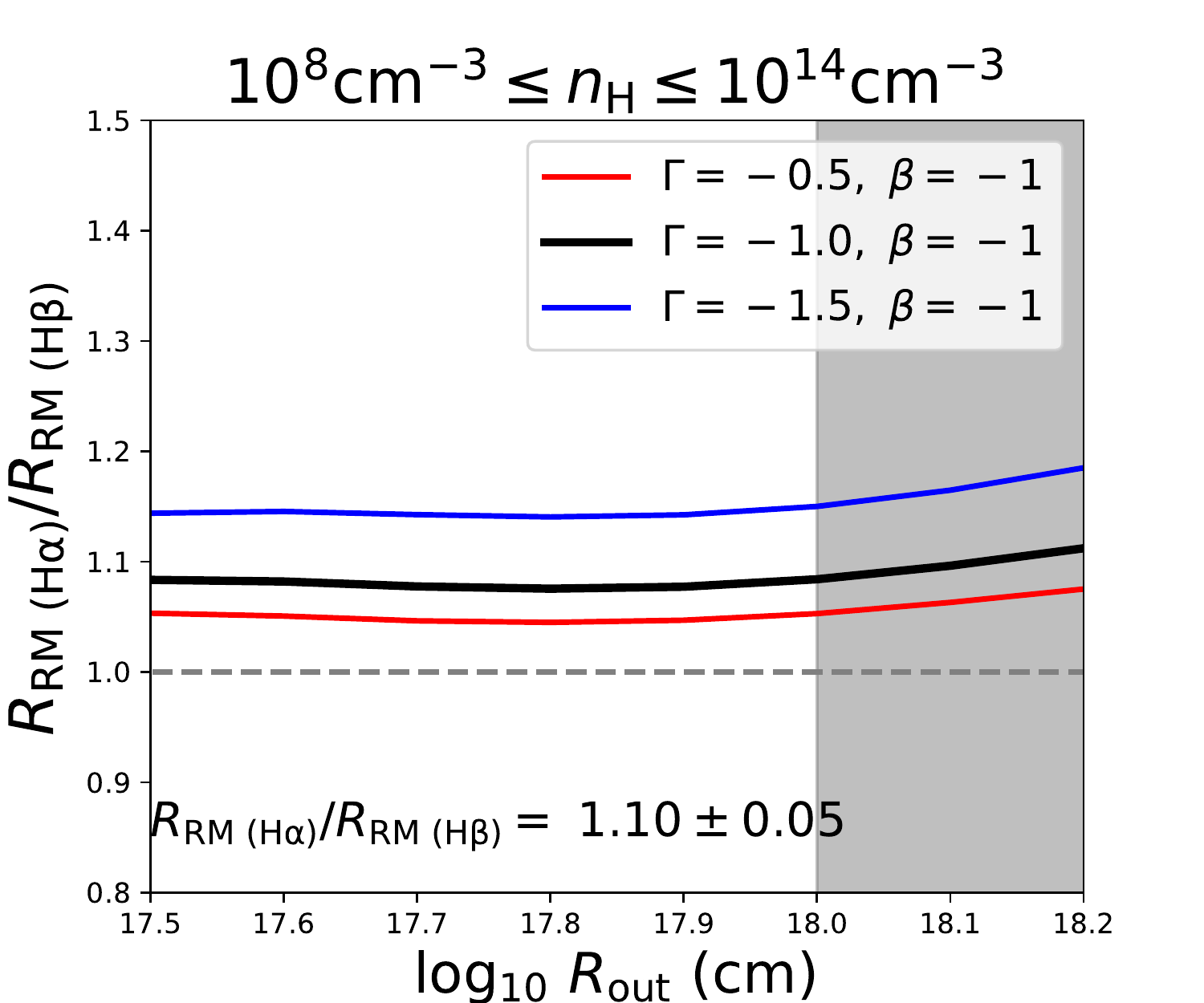}
	\caption{The simulation radius ratio of \ha\ to \hb\ of the RM measurements is $R_{\rm RM\ (\ha)}/R_{\rm RM\ (\hb)}=1.10\pm 0.05$.
All the parameters ranges and symbols are the same as Figure 4, except the gas density range. Here the density range is 
$10^8{\rm cm^{-3}} \leq n_{\rm H} \leq 10^{14} \rm cm^{-3}$. 
 }
\end{figure}

\section{The deviation between SA and RM size}
\label{sec:result}
\subsection{The simulation results}
We adopt a radius ratio $K=R_{\rm SA}/R_{\rm RM}$ to describe the deviation between of SA and RM size.
In this section, we will estimate the uncertainty of $K$ under the following parameter intervals which have been mentioned in the above section:
$ -1.5 <\Gamma < -0.5$, $\beta =-1$, $10^{8} \cc \le n_{\rm H} \le 10^{12}\cc$, $N_{\rm H}=10^{23}\cmii$ 
and $10^{17.5} {\rm cm} <R_{\rm out}<10^{18} {\rm cm}$.
We define the average and half-range of $K$ as: $\bar{K}=(K_{\rm max}+K_{\rm min})/2$ and $\Delta_K=(K_{\rm max}-K_{\rm min})/2$, 
where $K_{\rm max}$ and $K_{\rm min}$
are the maximum and minimum of $K$ under the parameter intervals, respectively.
We firstly calculate the $K$ for the combination of the same lines.
As shown in Figure \textbf{3a, 3b, 3c}, the average and half-range of $K$ are $0.859\pm 0.063$, $0.879\pm 0.067$, $0.851\pm 0.083$ 
for the pairs of \ha(SA)/\ha(RM), \pa(SA)/\pa(RM), and \hb(SA)/\hb(RM), respectively.
Surprisingly, all the averages of radius ratios $\Bar K$ obtained from these same lines are much lower than 1 ($\sim 0.85-0.88$).
For the combination of the different lines (Figure \textbf{3d, 3e, 3f}), the average and half-range of $K$ are $0.974\pm 0.042$, $0.947\pm 0.028$, $0.894\pm 0.061$ for the pairs of \pa(SA)/\hb(RM), \ha(SA)/\hb(RM), and \pa(SA)/\ha(RM), respectively.
Consequently, the corrected angular diameter distance $D^{\rm corr}_{\rm A} = R_{\rm SA}/\theta =K* R_{\rm RM}/\theta$=$K*D_{\rm A}$.
According to the $z-D_{A}$ relation \citep{peacock1998}, the corrected Hubble constant $H^{\rm corr}_0$=$H_0/K$, where $H_0$ is measured 
by the $D_{\rm A}$.

Interestingly, it clearly shows that not only the half-range of $K$ from the combination of the different lines are less than that from the 
combination of the same lines, but also the averages $\Bar K$ from the combination of the different lines are closer to 1. 
As a result, the combination of the different lines are more suitable for the SARM project than the combination of the same lines.
Among them, the $\Bar K$ values from the \pa(SA)/\hb(RM) and \ha(SA)/\hb(RM) pairs are closest to 1 and have the smallest uncertainty. 

At present, the interferometry technology has been achieved only in the infrared band K or 
longer wavelengths and for bright sources. For example, the GRAVITY of VLTI operates at 
the wavelengths range: 2.0–2.4 $\mu$m. For objects with a redshift $ z<0.28$ 
the \pa\ line is the only accessible strong broad line. Thus the \pa(SA)/\hb(RM) pair 
is the best choice for the low redshift object in the SARM project. In the future, with 
a much larger collection area of telescopes and the advanced technology, it is possible 
to conduct interferometric observations for faint sources and to shorter wavelengths,    
enabling SARM for intermediate redshift quasars. In that case, a combination of \ha(SA)/\hb(RM) 
would be fruitful.  

\subsection{Comparison between the simulation and observation for the RM measurements of \ha\ and \hb }
In order to test the reliability of our simulation, we compare the simulation with the observation for the RM measurements of \ha\ and \hb.
As shown in Figure 4\textbf{a}, the radius ratio of \ha\ to \hb\ of the RM measurements is $R_{\rm RM\ (\ha)}/R_{\rm RM\ (\hb)}=1.10\pm 0.04$.
The definitions of the average and half-range of $R_{\rm RM\ (\ha)}/R_{\rm RM\ (\hb)}$ are the same as for $K$.
The observational results from \citet{Kaspi2000}, \citet{Bentz2010} and \citet{Grier2017} are shown in Figure 4\textbf{b}. The black histogram in Figure 4\textbf{b} represents the sum of the above three papers.
The mean and standard deviation of the distribution of observational $R_{\rm RM\ (\ha)}/R_{\rm RM\ (\hb)}$ of the are $1.4\pm 0.5$.
Both this simulation and observational results suggest that the \ha\ region is slightly larger than that of \hb.
And our simulation result is roughly consistent with the observational results within a standard deviation.
Therefore, our parameters selection of LOC model and simulation results are in a reasonable range.

\subsection{Discussion}

In general, the range of the number density of BLR gas can be considered as:$10^{8} \cc \le n_{\rm H} \le 10^{12}\cc$. 
However, as shown in Figure 1, the locations of EW peaks of \ha, \hb\ and \pa\ lines are between $10^{12} \cc \le n_{\rm H} \le 10^{14}\cc$. 
Furthermore, the maximum value of BLR gas density can reach $10^{14}\cc$ under the radiation pressure confinement
(RPC, i.e., the equilibrium between the gas pressure and radiation pressure) model (\citealt{stern2014, baskin2014}). 
In view of this, it is necessary to check the simulation results when the maximum value of $n_{\rm H}$ reaches $10^{14} \cc$.
The radius ratios $K$ at the case of $10^{8} \cc \le n_{\rm H} \le 10^{14}\cc$ are shown in Figure 5. 
Similar to the case of $10^{8} \cc \le n_{\rm H} \le 10^{12}\cc$, not only the half-range of $K$ from the combination 
of the different lines (panels \textbf{d, e, f}) are less than that from the combination of the same lines (panels \textbf{a, b, c}), 
but also the averages of radius ratios $\Bar K$ from the combination of the different lines are closer to 1. 
Similar to Figure 3, the $K$ values from the \pa(SA)/\hb(RM) and \ha(SA)/\hb(RM) pairs are also closest to 1 and have the smallest 
uncertainty. Furthermore, as shown in Figure 6, the radius ratio of \ha\ to \hb\ of the RM measurements is 
$R_{\rm RM\ (\ha)}/R_{\rm RM\ (\hb)}=1.10\pm 0.05$, which is similar to the case of $10^{8} \cc \le n_{\rm H} \le 10^{12}\cc$
(Figure 4). As a result, no matter the maximum value of the number density of BLR gas $10^{12} \cc$ or $10^{14} \cc$, 
the \pa(SA)/\hb(RM) and \ha(SA)/\hb(RM) pairs are always the best two choices for the SARM project. 

As mentioned above, the RPC model \citep{baskin2014} is a more physically realistic BLR model which assumes that the gas pressure and radiation pressure are in equilibrium. In the RPC model, the cloud is no longer a uniform density slab, but has a density structure.
In our next work, we will calculate the deviation between SA and RM size under the RPC model and make a comprehensive and detailed 
comparison with the LOC model.

\section{Conclusion}
\label{sec:con}
The combination of interferometric and reverberation mapping for BLR of AGNs, can be used to measure the Hubble 
constant $H_0$. However, there may be a systematic deviation between these two different radii $R_{\rm SA}$ and $R_{\rm RM}$. 
In this work, we calculate this systematic deviation for three hydrogen lines \ha, \hb, \pa\ based on the LOC model of BLR. 
We estimate the half-range of radius ratio $K=R_{\rm SA}/R_{\rm RM}$ under sufficiently wide parameter ranges:
$ -1.5 <\Gamma < -0.5$, $\beta =-1$, $10^{8} \cc \le n_{\rm H} \le 10^{12}\cc$, $N_{\rm H}=10^{23}\cmii$ 
and $10^{17.5} {\rm cm} <R_{\rm out}<10^{18} {\rm cm}$.
Our main results can be summarized as follows:
\begin{itemize}
\item[1.]  The ratios for the same line are systematically lower than unity by 10-15\% for 
all hydrogen lines considered here (i.e.,\ha(SA)/\ha(RM), \pa(SA)/\pa(RM), and \hb(SA)/\hb(RM) ) and 
the scatter in these ratios are typically 0.06-0.08.  

\item[2.] The $K$ values from the \pa(SA)/\hb(RM) and \ha(SA)/\hb(RM) pairs are closest to 1 and have 
the smallest uncertainty.
\end{itemize}

Considering the current infrared interferometry technology, the \pa(SA)/\hb(RM) pair is the best 
choice for the low redshift object in the SARM project. In the future, the \ha(SA)/\hb(RM) pair could be used 
for the high redshift luminous quasars in the SARM project.

\acknowledgments
We thank Prof. Jian-min Wang for the helpful discussion and comments on this work.
We thank the anonymous referee for valuable comments and constructive suggestions.
Z.-C. H. is supported by NSFC-11903031 and USTC Research Funds of the Double First-Class Initiative YD 3440002001.
T.-G. W. is supported by NSFC-11833007. H. -X. G. acknowledges the NSF grant AST-1907290.

\bibliography{rmsa}{}

\end{document}